\RequirePackage{fix-cm}
\documentclass[twocolumn]{svjour3}          % twocolumn
\smartqed  % flush right qed marks, e.g. at end of proof

\usepackage{graphicx}

\usepackage{amssymb}
\usepackage{eqnarray,amsmath}

\begin{document}

\title{Experiments on barotropic-baroclinic conversion and the applicability of linear $n$-layer internal wave theories}


\author{Mikl\'os Vincze        \and
        Tam\'as Boz\'oki %etc.
}


\institute{M. Vincze \at
              MTA-ELTE Theoretical Physics Research Group, P\'azm\'any P. stny. 1/a, H-1117 Budapest, Hungary \\
              Tel.: +36-70-3103352
              \email{mvincze@general.elte.hu}           %  \\
           \and
           T. Boz\'{o}ki \at
           von K\'arm\'an Laboratory for Environmental Flows, P\'azm\'any P. stny. 1/a, H-1117 Budapest, Hungary               
}

\date{Received: date / Accepted: date}

\maketitle

\begin{abstract}
Interfacial internal waves in a stratified fluid excited by periodic free-surface perturbations {in a closed tank} are studied experimentally. {Barotropic-baroclinic energy conversion is induced by} 
the presence of a bottom obstacle. The connection between horizontal surface velocities and internal wave amplitudes is investigated, the developing flow patterns are described qualitatively, and the wave speeds of internal waves are systematically analyzed and compared to linear 2- and 3-layer theories. We find that, despite of the fact that the observed internal waves can have considerable amplitudes, a linear 3-layer approximation still gives fairly good agreement with the experimental results. 
\end{abstract}

\section{Introduction}
\label{intro}

Internal gravity waves play an essential role in the dynamics of natural water bodies ranging from stratified lakes \cite{lake_internal} to oceans \cite{gm,alapmu}. Energy and momentum transfer between surface waves and their internal counterparts in the bulk is ubiquitous in nature: exchange flows between differently stratified connected basins \cite{knight_inlet}, tidal conversion at seafloor sills in the deep ocean \cite{tidal_conversion,kelly,holloway}, or the so-called dead water effect that converts the kinetic energy of a moving ship to interfacial wave energy in stratified fjords \cite{deadwater} are just a few examples of its occurrence on different scales.    

Coastal areas with river (or glacier) runoff or regions in the open ocean with steep thermocline \cite{two_layer,north_sea}
can be treated as a system of uniform or linearly stratified water layers located stably above each other.
Waves propagating along the interfaces between the layers can be either barotropic or baroclinic in nature. In the former case the vertical oscillations of the considered interface is in phase with that of the water surface and their amplitudes and wave speeds are of the same order of magnitude. In baroclinic waves, however, the internal waves may exhibit large amplitudes without practically any noticeable displacement at the surface.

The above framework of dividing continuously stratified vertical density profiles to discrete layers is referred to as $n$-layer approach. 
The classic linear 2-layer theory assumes homogeneous layers of constant densities, rigid bottom, irrotational flow, and small-amplitude disturbances at the interface and at the free surface. This approximation is valid as long as the considered density difference is small compared to the reference density and the thickness of the pycnocline (i.e. the `gradient zone' between the layers) is small relative to the total depth and the wavelength. A thorough introduction can be found in the work of Sutherland \cite{Sutherland}.
For long waves of larger amplitudes weakly nonlinear 2-layer theories are available based on the periodic solutions of the 2-layer Korteweg--de Vries (KdV) equation \cite{cnoid,kundu,alapmu2}, whereas for shorter waves other nonlinear models exist, e.g. \cite{Hunt1961}. 3-layer approach enables a more advanced treatment of the dynamics. Here we refer to the fully nonlinear analysis of Fructus and Grue \cite{FG}: their approximation is based on piecewise-linear stratification profiles (a setting motivated by internal solitary wave experiments), and as a byproduct, they also provide a linear framework for relatively small-amplitude waves.    

The energy of barotropic flow can be converted to excite baroclinic wave modes in the bulk by the interaction of flow and topography.  
In an earlier study from our laboratory \cite{kozma} it has been demonstrated that a single thin vertical obstacle placed to the bottom in the middle of a rectangular tank excites such interfacial waves if a small-amplitude barotropic standing wave ({seiche}) is
present on the water surface.  
A follow-up paper \cite{boschan} investigated resonant barotropic-baroclinic conversion in the presence of two identical thin obstacles. 
Here we further investigate the coupling between surface waves and interfacial wave propagation and test the applicability of the two- and 3-layer linear wave theories.   
  
The paper is organized as follows. Section 2 describes the experimental set-up and briefly introduces the applied methods. We present our results and their comparison with 2- and 3-layer theories in Section 3, and further discuss our findings and give brief conclusions in Section 4. 

\section{Experimental set-up and measurement methods}
\label{setup}
\begin{figure}[b!]
\noindent\includegraphics[width=\columnwidth]{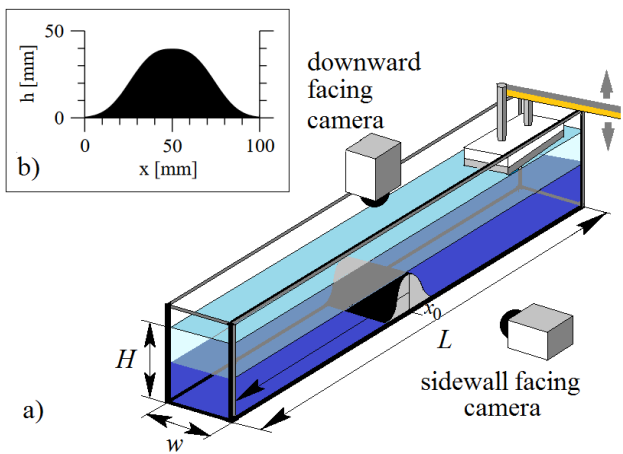}
\caption{(a) The experimental tank and the two camera positions. The geometrical parameters of the tank are $L=239$ cm, $w = 8.8$ cm, and $H = 9$ cm. (b) The cross section of the bottom obstacle used: a 4 cm-tall Gaussian with 4.24 cm full width at half maximum.}
\label{setup}
\end{figure}
Our experiments were conducted in a rectangular acrylic tank of length $L=239$ cm and width $w = 8.8$ cm, filled up to level $H = 9$ cm with quasi-2-layer stratified water (Fig.\ref{setup}a). The bottom layer consisted of saline water, whereas the upper layer was formed by pure tap water of the same temperature (approximately $24^\circ$C). {Four} measurement series were carried out -- hereafter referred to as series \#1, \#2, \#3, {and \#4} -- with different vertical density profiles, shown in Fig.\ref{profiles}. 

A $h_0 = 4$ cm-tall obstacle, a prism of Gaussian cross section (Fig{.\ref{setup}b) with 4.24 cm {full width at half maximum} was placed at the bottom. The top of the obstacle in each case was located {at horizontal position $x_0 = 110$ cm ($ = 0.46 L$), see Fig{.\ref{setup}a.} 
{This slight offset from the center was chosen in order to avoid being exactly at the antinode of certain seiching modes of the water surface, where the horizontal flow would practically vanish when excited with the corresponding eigenfrequencies (see Subsection 3.2).}
Waves were excited on the free water surface with a characteristic amplitude of 0.5 cm by a 15 cm long and 6 cm wide wave maker made of foam rubber, as sketched in Fig.\ref{setup}a. The wave maker was oscillating vertically, driven by a rotating DC motor whose angular frequency was adjusted between $\omega=0.5$ rad s$^{-1}$ and $\omega=7$ rad s$^{-1}$.

In series \#1, \#2 {and \#3 blue dye was dissolved in the bottom layer and} each experiment was recorded with a video camera (at frame rate 30 fps and frame size $480 \rm{px} \times 640\rm{px}$) pointing perpendicularly to the sidewall close to the middle of the tank. \emph{Tracker}, an open source correlation based pattern tracking software \cite{tracker} was used to obtain time series of the vertical motion of the water surface and the relatively sharp interface between the saline (blue) and pure (transparent) layers at certain horizontal locations.

{The density profiles of the water body in the different series are shown in Fig.\ref{profiles}. As the present work focuses firstly on the 2-layer approximation and its applicability, we calculated the `effective thickness' of the layers by fitting the function form $\rho(z)=K\tanh(B\cdot(z-H^{(2)}_2))+D$ -- shown with dashed curves in Fig.\ref{profiles} -- and taking $H^{(2)}_2$ as the unperturbed thickness of the bottom layer. (Hereafter the upper index in parenthesis -- 2 or 3 -- denotes whether two- or three- layer approximation is applied.) Top layer thickness $H^{(2)}_1$ was then assigned to be $H^{(2)}_1 = H-H^{(2)}_2$, where $H=9$ cm is the total water depth in all cases. The `effective density' $\rho_2$ of the bottom layer was then obtained from the fitted function as $\rho_2=1/H^{(2)}_2\int^{z=H^{(2)}_2}_{z=0}\rho(z)dz$, and in similar manner for the top layer: $\rho_1=1/H^{(2)}_1\int^{z=H}_{z=H^{(2)}_2}\rho(z)dz$, as proposed, e.g. in \cite{CT}. 

\begin{table*}[t]
\begin{center}
    \begin{tabular}{| c | c | c | c | c | c | c |}
    \hline
    Exp. ser. & $H^{(2)}_1$ [cm] & $H^{(2)}_2$[cm] & $H_r$ [cm] & $\rho_1$ [kg/l] & $\rho_2$ [kg/l] & $c^{(2)}_0$ [cm/s] \\ \hline
      \#1 & 5.0 & 4.0 & 2.22 & 1.001 & 1.069 & 12.10 \\ \hline
      \#2 & 4.3 & 4.7 & 2.25 & 1.001 & 1.067 & 11.84 \\ \hline
      \#3 & 5.4	& 3.6 & 2.16 & 1.001 & 1.059 & 11.09 \\ \hline
    \end{tabular}
\caption{The geometrical and material properties of the set-up for experiment series \#1, \#2, and \#3 in the 2-layer framework.} 
\label{table}
\end{center}
\end{table*}
\begin {table*}
\begin{center}
    \begin{tabular}{| c | c | c | c | c | c | c | c |}
    \hline
    Exp. ser. & $H^{(3)}_1$ [cm] & $H^{(3)}_2$ [cm] & $H^{(3)}_3$ [cm] & $N_1$ [rad s$^{-1}$] & $N_2$ [rad s$^{-1}$] & $N_3$ [rad s$^{-1}$] & $c^{(3)}_0$ [cm/s] \\ \hline
      \#1 & 4.47 & 1.15 & 3.38 & 0.41 & 7.59 & 0.48 & 11.62 \\ \hline
      \#2 & 3.37 & 2.03 & 3.60 & 0.70 & 5.66 & 1.20 & 11.37 \\ \hline
      \#3 & 4.84 & 0.997 & 3.16 & 0.97 & 7.15 & 1.90 & 10.54 \\ \hline
    \end{tabular}
\caption{The layer thicknesses, buoyancy frequencies (\ref{N}) and the corresponding long-wave speeds in the 3-layer framework for experiment series \#1, \#2, and \#3.} 
\label{table2}
\end{center}
\end {table*}
\begin{figure}[]
\noindent\includegraphics[width=\columnwidth]{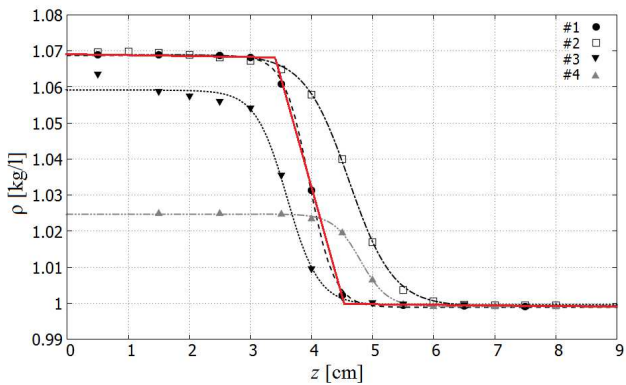}
\caption{Density $\rho$ as a function of vertical position $z$ from the bottom for all {four} experiment series (see legend), as acquired with a conductivity probe. Their tanh fits are also shown. The red solid curve shows the 3-layer piece-wise linear fit to the profile of experiment series \#1 used for the 3-layer approximation.}
\label{profiles}
\end{figure}

{In the 3-layer treatment, the values of buoyancy-, or Brunt-V\"ais\"al\"a frequencies $N_j$ are also calculated for each layer $j$. The value of $N_j$ is obtained as
\begin{equation}
N_j\equiv \sqrt {-\frac{g}{\rho_0} \frac{d\rho}{dz}\Big|_j},
\label{N}
\end{equation}
where $\rho_0=1$ kg dm$^{-3}$ denotes the reference density, $g$ is the gravitational acceleration, and subscript $j$ denotes that the domain is restricted to layer $j$. The derivatives can be estimated from the slopes of the density profiles via piece-wise linear fits to the different sections of the measured $\rho(z)$. The $z$-coordinates of their intersection points yield the desired values of layer thicknesses $H^{(3)}_j$. The procedure is demonstrated on the profile of experiment series \#1 in Fig.\ref{profiles} (red curve). The parameters obtained this way for series \#1, \#2, and \#3 are summarized in Table \ref{table2}.}

Experiment series \#4 was conducted in order to gather supplementary information on the velocity fields in the set-up. We analyzed the flow velocities {in a horizontal plane close to} the water surface;
the camera was therefore installed in a downward facing position (Fig.\ref{setup}a) above the obstacle 
{with its field of view centered to $x_0$ and covering a 10 cm-wide domain}.
The water surface was seeded with particle image velocimetry (PIV) tracers (of diameter $\sim 100\rm{\mu m}$) whose patterns were tracked with the Tracker software.  In series \#4 the prepared salinity of the bottom layer was lower than in the other two profiles (see Fig.\ref{profiles}) in order to make data acquisition with PIV technique \cite{pivpaper} possible; thus the fluid density matched better that of the PIV tracers, which otherwise could not have been {distributed} uniformly in the working fluid. {Beside the series of the horizontal surface velocity measurements} we conducted PIV measurements in the vertical plane as well -- from the sidewall facing camera position -- in order to gain a qualitative impression of the internal flow field.

\section{Results}
\label{results}
\subsection{General description of the flow}
\begin{figure*}[]
\centering
\noindent\includegraphics[width=0.85\textwidth]{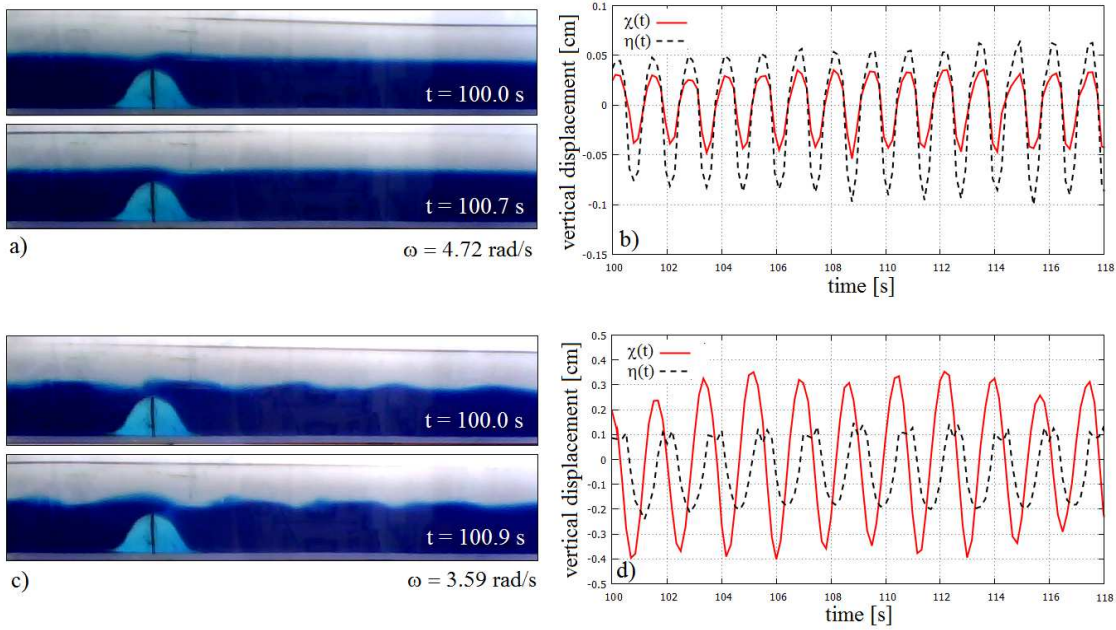}
\caption{Barotropic and baroclinic modes.(a) Two snapshots of a barotropic wave mode ($\omega=4.72$ rad s$^{-1}$), and the corresponding vertical displacement time series (b) at the surface, $\eta(t)$ (dashed) and at the interface, $\chi(t)$ (red solid curve) 43.5 cm to the right of the obstacle. (c) Two snapshots of a baroclinic wave mode ($\omega=3.59$ rad s$^{-1}$), and the corresponding time series (d), as in the previous case. Note that the vertical scales in panels (b) and (d) are different.}
\label{snapshots}
\end{figure*}

\begin{figure*}[]
\centering
\noindent\includegraphics[width=0.85\textwidth]{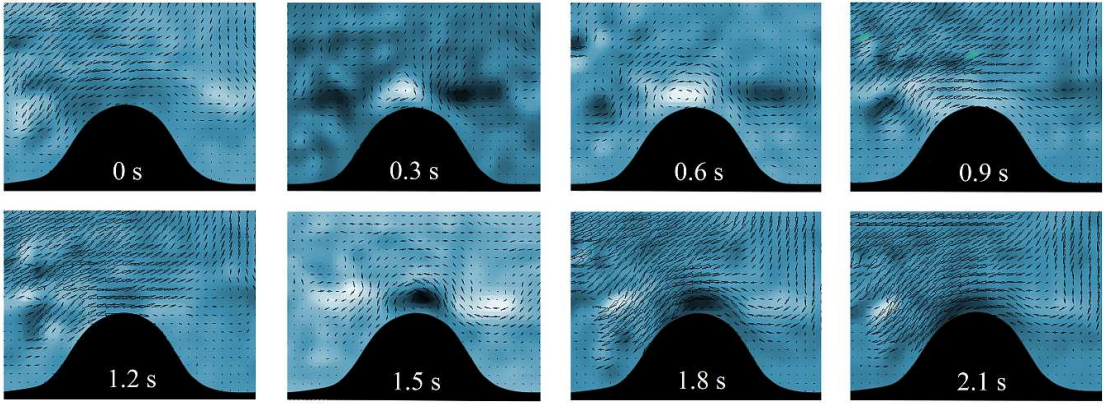}
\caption{PIV snapshots at forcing frequency $\omega=2.58$ rad s$^{-1}$ at the vicinity of the obstacle. The time labels indicate the time relative to the first image. The shading denotes the vorticity field: light (dark) areas represent clockwise (counterclockwise) flow.}
\label{piv}
\end{figure*}

The wave dynamics in the set-up is driven by the time-dependent horizontal flow above the obstacle. This is imposed on the water surface by the vertical motion of the wave maker oscillating with the same amplitude in all experiments. 
{For forcing frequencies above a certain threshold $\omega_*$ the oscillations of the interface are \emph{barotropic}: snapshots from such a typical experimental run ($\omega=4.72$ rad s$^{-1}$) are shown in Fig.\ref{snapshots}a.    
In this case the flow of the bottom layer does not have enough time to pass over the obstacle and so the vertical `sloshing' of the interface $\chi$ remains localized. Farther away from the obstacle the internal vertical oscillation is in phase with that of the water surface $\eta$ with larger amplitudes at the surface, as demonstrated in Fig. \ref{snapshots}b. Here the simultaneous time series of $\eta(t)$ (dashed line) and $\chi(t)$ (solid line) are shown at the same horizontal position (at a horizontal distance of 43.5 cm -- i.e. more than $10 h_0$ -- from the obstacle).}

{In the baroclinic regime fluid parcels from the bottom layer are transported above the top of the obstacle by the flow in each period and their run-up yields the formation of billows connected to the obstacle as visible in Fig.\ref{snapshots}c ($\omega=3.59$ rad s$^{-1}$). These billows are formed due to shear instability and can reach and penetrate the interface on the other side of the obstacle and thus initiate \emph{baroclinic} internal waves which radiate away from the obstacle in both horizontal directions. { (The precise classification of the instability generating the billows is beyond the scope of the present study.)} The corresponding time series of vertical displacement -- from the same location as in the previous case -- are presented in Fig. \ref{snapshots}d. The difference in the amplitudes (larger at the interface) and phases is clearly visible.}

The formation of localized billows at the top of the obstacle is qualitatively demonstrated with the vertical-plane PIV snapshot sequence of Fig.\ref{piv} obtained in series \#4, at forcing frequency $\omega=2.58$ rad s$^{-1}$. The time values shown in the images represent the elapsed time from the first snapshot. The darkness scale in the background of the velocity vector field marks vorticity; light (dark) areas denote clockwise (counterclockwise) flow. As expected, the strongest vorticity appears at the top of the obstacle (cf. Fig.\ref{snapshots}c). First at around 0.3 s, corresponding to leftward flow in the upper layer and then in counterphase, i.e. half a period (around 1.2 s) later, induced by rightward flow. 

\subsection{The frequency-dependence of the barotropic-baroclinic conversion}
\label{forcing}
\begin{figure*}[]
\centering
\noindent\includegraphics[width=0.85\textwidth]{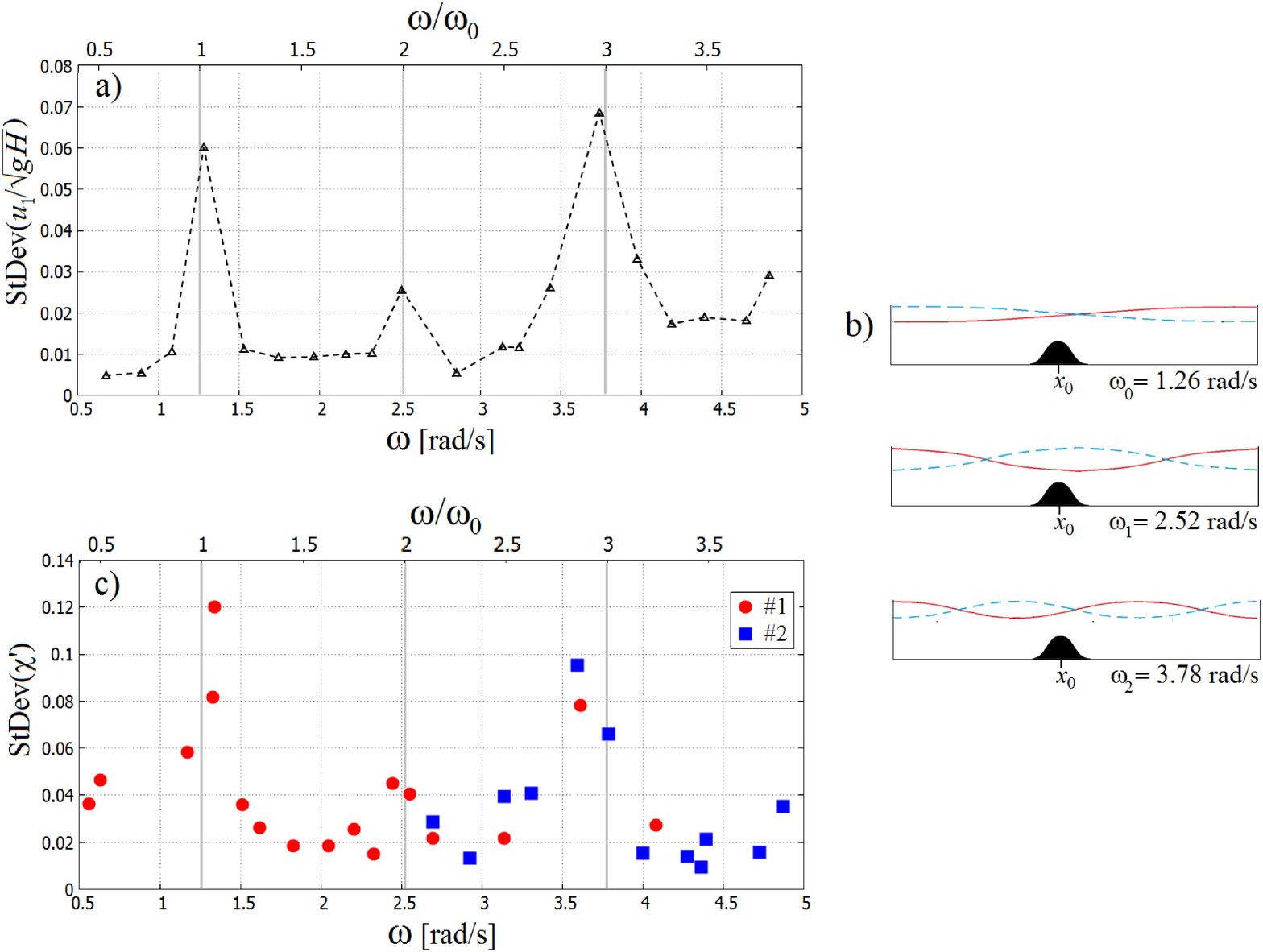}
\caption{(a) The standard deviations of the \emph{horizontal} velocity time series (in $\sqrt{gH}$ units) above the obstacle as a function of forcing angular frequency $\omega$ based on particle tracking from series \#4. (b) Cartoons of the barotropic standing wave (seiche) modes with their theoretical frequency. The wave forms marked by the red solid and blue dashed curves denote the two extrema of wave displacement, half a period apart. Amplitudes are not drawn to scale. (c) The standard deviations of the nondimensional \emph{vertical} interface displacement time series, acquired 4 cm to the right from the top of the obstacle for series \#1 (red circles) and \#2 (blue squares).} 
\label{tracersfromabove}
\end{figure*}

{Although the wave maker was oscillating with the same amplitude in all experiments, the displacement of the free water surface exhibited strong dependence on the frequency of the forcing.} To quantitatively address this connection the characteristic horizontal velocity in the top layer $u_1(x_0,t)$ at the obstacle location $x_0$ 
we tracked PIV tracers in the horizontal plane about 0.5-1 cm below the free surface (from the downward pointing camera position, see Fig.\ref{setup}) and averaged the velocity vectors in the field of view. Thus a field-average of the horizontal velocity $\langle u_1 \rangle$ in the vicinity of $x_0$ was determined at each time instant $t$ with an acquisition rate of 6 fps.       
The standard deviations of this time series ${\rm StDev}(u_1)=\sqrt{\langle u_1(t)^2 \rangle - \langle u_1(t) \rangle^2}$ obtained for $\mathcal{O}(100)$ forcing periods in each case are shown in Fig.\ref{tracersfromabove}a in the units of the shallow layer barotropic wave speed $\sqrt{gH}$ as a function of the imposed angular frequency $\omega$. The amplification of velocities at the fundamental surface seiche frequency $\omega_0$ and its harmonics is clearly visible in good agreement with the shallow-layer formula 
\begin{equation}
\omega_n=(n+1)\frac{2\pi\sqrt{gH}}{2L}
\end{equation}
with $n=0,1,2,\dots$ being the mode index, yielding $\omega_0=1.26$ rad s$^{-1}$, $\omega_1=2.52$ rad s$^{-1}$, and $\omega_2=3.78$ rad s$^{-1}$ \cite{Sutherland}. We note that the forcing frequency is also expressed in nondimensional units $\omega/\omega_0$ on the top horizontal axes of panely Fig.\ref{tracersfromabove}a and c.

{The amplitude of the internal waves is set by the characteristic size of the aforementioned billows and is therefore determined by the interfacial shear in the vicinity of the obstacle. 
Here the horizontal flow in the bottom layer is partially blocked ($u_2(x_0,t) \approx 0$), thus the velocity difference at the interface can be approximated with $u_1(x_0,t)$. Therefore, one would expect a strong correlation between the amplitude of the oscillating current above the obstacle and the observed internal wave amplitudes.}

Fig.\ref{tracersfromabove}c shows the standard deviations of the nondimensional vertical interface displacement time series {$\chi'(t)$, expressed relative to the unit characterinstic scale $H_r = H^{(2)}_1 H^{(2)}_2/H$, measured close to the obstacle's location (4 cm to the right from $x_0$) for the investigated frequencies in series \#1 (red circles) and \#2 (blue squares). The standard deviations were calculated in the same manner as ${\rm StDev}(u_1)$ for panel a.}  The correlation of horizontal surface velocities and the vertical wave amplitudes is apparent.

{The differences in the size of the resonant peaks in Fig.\ref{tracersfromabove}a can be explained by the following reasoning. The horizontal velocity field $u_1(x,t)$ in the upper layer is closely connected to the phase and amplitude of the surface wave above it. According to the linear one-dimensional theory \cite{Sutherland} $u_1(x,t)$ and surface height $\eta(x,t)$ at position $x$ and time $t$ are related as
\begin{equation}
\frac{\partial u_1}{\partial t}=-g\frac{\partial \eta}{\partial x}.
\label{eta}
\end{equation}  
}

{Solving equation (\ref{eta}) for the standing wave forms $\eta=\eta_0 \cos(x (n+1)\pi/L) \cos(t (n+1)\sqrt{gH}\pi/L)$ shown in the sketches of Fig.\ref{tracersfromabove}b and calculating the standard deviations of the time series obtained from the sinusoidal solution $u_1 (x_0,t)$ at position $x_0$ of the obstacle, one gets a 0.26 times smaller value for the peak of the first harmonic ($n=1$) than for the fundamental mode. Compared to this, the observed value of this ratio from the experiment is somewhat larger, around 0.42 (cf. Fig.\ref{tracersfromabove}a).}

As a secondary effect, barotropic-baroclinic energy  conversion provides a certain damping of the seiching modes at the surface, contributing to the sizes of the resonant peaks. 
%According to the theory \cite{lake_internal}, barotropic wave energy would decay exponentially as $E(t) \propto \exp(-Ct)$ if the barotropic forcing would cease at time $t = 0$. The damping coefficient $C$ scales with the square of the amplitude $U$ of oscillating horizontal current, $C \propto U^2(x_0)$, as measured at the location $x_0$ of the obstacle. At the seiching eigenmodes $U(n,x0) \propto \sin(x_0(n+1)\pi/L)$ holds. Substituting the location of the obstacle ($x_0/L = 0.46$) one gets $U^2(0,x0)/U^2(2,x0) \approx 1.14$, implying that the damping of the surface seiche due to barotropic-baroclinic energy conversion is stronger by this factor in the case of the fundamental mode $n = 0$ than for $n = 2$. This is in good qualitative agreement with the data points in Fig.\ref{tracersfromabove}a and c: 
The surface current above the obstacle is somewhat smaller for the $n=0$ fundamental mode than that for harmonic $n=2$ (panel a), whereas the internal wave amplitudes are larger for $n=0$ (panel c). This implies that the damping of the surface seiche due to baroclinic wave excitation is stronger in the case of the fundamental mode $n = 0$ than for $n = 2$.
  
\subsection{Internal wave characteristics}
In our experiments four types of wave propagation can be observed on the interface, examples of which are shown in the space-time plots of Fig.\ref{hovmoellers} corresponding to different forcing frequencies $\omega$.
The position $x_0$ of the obstacle is marked with a black triangle in each panel.
The coloring of a data point at $(x,t)$ represents the integrated total darkness of the vertical pixel column at horizontal position $x$ acquired from the video frame of the sidewall-pointing camera at time $t$ (like the ones in Fig. \ref{snapshots}a and c) and is therefore proportional to the vertical displacement of the interface. The coloring in each panel is normalized with respect to the largest and smallest values in the given space-time plot for better visibility, therefore the amplitudes of the different cases cannot be compared to each other.    

\begin{figure}[]
\noindent\includegraphics[width=\columnwidth]{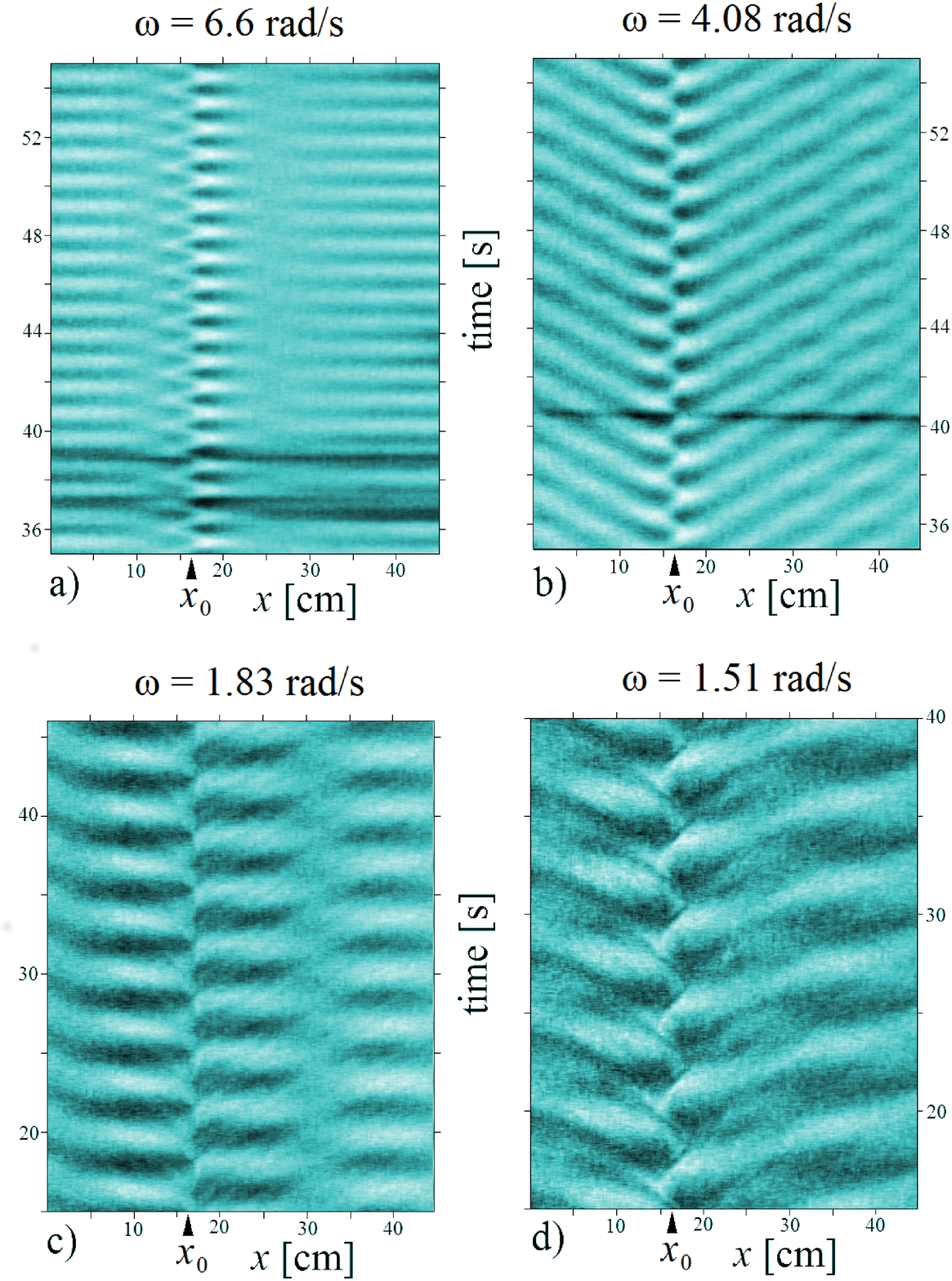}
\caption{Space-time plots of four experimental runs, demonstrating typical wave patterns. (a) Barotropic oscillation at $\omega = 6.6$ rad s$^{-1}$. (b) Baroclinic short-wave propagation at $\omega = 4.08$ rad s$^{-1}$. (c) Interfacial standing wave at $\omega = 1.83$ rad s$^{-1}$. (d) Modulated propagation of long interfacial wave patterns at $\omega = 1.51$ rad s$^{-1}$. The position of the obstacle $x_0$ is marked with black triangles.}
\label{hovmoellers}
\end{figure}

Fig.\ref{hovmoellers}a shows a barotropic oscillation at forcing frequency $\omega = 6.6$ rad s$^{-1}$. Note that -- in agreement with Fig. \ref{snapshots}a -- in this particular case the interface levels on the left and right hand sides of the obstacle are in counterphase and no internal wave excitation can be observed.
In Fig.\ref{hovmoellers}b at $\omega = 4.08$ rad s$^{-1}$ the propagation of small-wavelength baroclinic internal waves is clearly visible. In this case wave reflection from the vertical endwalls of the tank is negligible; these small-scale waves are damped by the barotropic counterflow in the upper layer and other viscous effects, therefore, by the time they reach the end of the basin (out of the horizontal range of the space-time plots) their amplitudes decay significantly. For lower frequencies, however, wave reflection becomes dominant: with increasing wavelength damping decreases and the superposition of the incident and reflected waves yields internal standing waves, as the ones seen in Fig.\ref{hovmoellers}c for $\omega = 1.83$ rad s$^{-1}$.

For even lower frequencies the typical space-time plots show a certain `meandering' (acceleration, followed by deceleration of the propagating structures), as depicted in panel d) for $\omega = 1.51$ rad s$^{-1}$. This modulation of wave speed can be interpreted as an interference between incident and reflected waves with the latter having smaller (but not anymore negligible) amplitudes. For a qualitative understanding of the effect let us consider two sinusoidal plane wave components, one traveling rightwards as $\chi_i = A_i \sin(-kx+\omega t)$ with amplitude $A_i$, frequency $\omega$ and wave number $k$ and another moving in the opposite direction with the same phase parameters but a smaller amplitude $A_r$ as $\chi_r = A_r \sin(kx+\omega t)$. Then their superpostion $\chi_i+\chi_r$ yields modulated propagation with an average speed equal to phase velocity $\omega/k$, as sketched in Fig.\ref{wavesketch}. (Note, that the classic standing wave state corresponds to $A_i=A_r$, \cite{kundu}.)

\begin{figure}
\centering
\noindent\includegraphics[width=0.8\columnwidth]{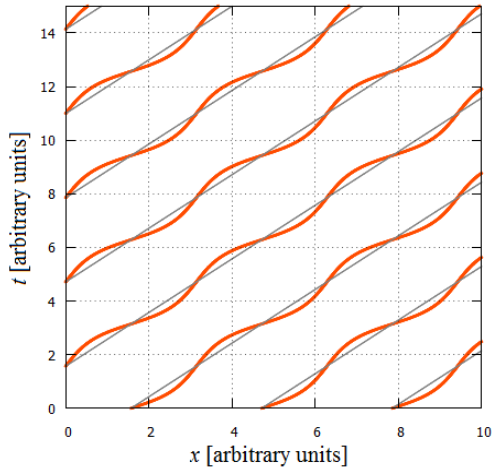}
\caption{A conceptual demonstration of the `meandering' propagation with sinusoidal waves. The propagation of the local maxima and minima of the $A_i \sin(-x+t)+A_r \sin(x+t)$ wave superposition is shown in case of $A_i = 1.5$ with $A_r=0.5$ (thick red curves) and without reflection, i.e. $A_i=1.5$ and $A_r=0$ (thin gray lines).}
\label{wavesketch}
\end{figure}

\subsection{2-layer approximation}
As length scale for the presentation of the results in the 2-layer framework in order to obtain data collapse, we introduced reduced height $H_r$ defined already in subsection 3.2 as the harmonic mean of $H^{(2)}_1$ and $H^{(2)}_2$, i.e. $H_r=H^{(2)}_1 H^{(2)}_2/H$ for each experiment series. The velocity values were rescaled with the theoretical baroclinic long-wave velocity of the 2-layer approximation $c^{(2)}_0 = \sqrt{g\,H_r(\rho_2-\rho_1)/\rho_0}$, see e.g. \cite{alapmu}.
As follows from the above relations, the unit of time characteristic to internal wave dynamics becomes $H_r/c^{(2)}_0$. Formally, the transformations of the relevant quantities are as follows: $k'=k H_r$, $\omega'=\omega H_r/c^{(2)}_0$, and $c'=c/c^{(2)}_0$ are the nondimensional versions of wavenumber $k$, frequency $\omega$ and wave speed $c$, respectively. The actual values of the scaling parameters for the four settings are summarized in Table \ref{table}.

\begin{figure*}[]
\noindent\includegraphics[width=\textwidth]{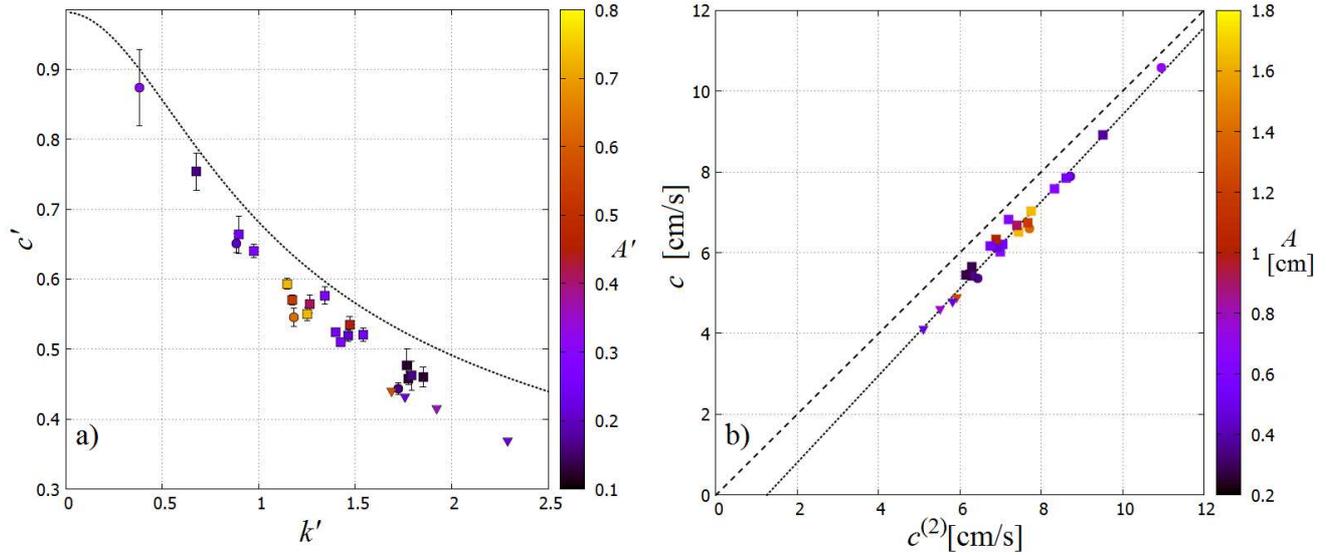}
\caption{(a) Nondimensional wave speeds $c'$ as a function of wave number $k'$ for series \#1 (circles), series \#2 (squares), and series \#3 (triangles). The coloring is based on the nondimensional peak-to-trough amplitude $A'$ in panel.
(b) Scatter plot of the theoretical values of 2-layer wave speed $c^{(2)}$ corresponding to the applied forcing frequencies (horizontal axis) and the measured values $c$ (vertical axis) in dimensional units. The dashed curve represents $c=c^{(2)}$, whereas the dotted line represents the linear fit to the data. Symbol coding is the same as in panel (a), the color bar here represents dimensional peak-to-trough amplitudes.}
\label{c_k_1}
\end{figure*}

Applying this non-dimensionalization the `empirical speeds' of the baroclinic waves $c'$ were evaluated by taking the slopes of linear fits to the iso-grayness contours of the respective space-time plot in the domain to the right of the obstacle. The `empirical wavenumber' $k'$ was than determined as $\omega'/c'$ using the known dimensionless excitation frequency $\omega'$. For this analysis only the `freely propagating' baroclinic waves (of the type shown in \ref{hovmoellers}b) were considered.

The observed velocities $c'$ as a function of wavenumber $k'$ are shown in Fig.\ref{c_k_1}a. Data from experiment series \#1, \#2, and \#3 are marked with circles, squares and triangles, respectively. The error bars represent the residual standard deviations of the aforementioned fitting procedures (for the data points where no error bars are visible the residuals were smaller than the size of the data point itself). The coloring of the data points indicates the nondimensional amplitudes $A'=A/H_r$ of the waves, where amplitude $A$ was determined as the maximum peak-to-trough vertical displacement of the interface in the given experiment.    

Also shown in panel a) is the theoretical $k'$-dependence (dashed line) of the phase velocity of small-amplitude (linear) waves propagating on the inner interface of an incompressible and irrotational 2-layer fluid with rigid top surface \cite{two_layer}. The formula reads as:
\begin{equation}
c^{(2)}=\frac{\omega}{k}=\sqrt{\frac{g}{k}\frac{\rho_2-\rho_1}{\rho_1\coth(H^{(2)}_1k)+\rho_2\coth(H^{(2)}_2k)}},
\label{linear_c_k}
\end{equation}   
where all notations are as before. {Transforming the above formula into the aforementioned nondimensional units yields the same curve for all considered experiment series.}

Panel b) demonstrates the correlation plot between the theoretical $c^{(2)}$ and the measured values, this time in dimensional units (the coloring of the data points, indicating the largest peak-to-trough amplitudes is also dimensional). The dashed curve marks $c = c^{(2)}$. Apparently, the 2-layer theory systematically overestimates the wave speeds in the set-up. The dotted line shows the linear fit to all points, yielding $c=1.08 c^{(2)} - 1.35$ cm/s. As expected, the theory performs better at higher velocities, corresponding to smaller wavenumbers: if the wavelength is significantly larger than the pycnocline thickness 3-layer corrections are negligible. 

\subsection{3-layer approximation}

{In small-amplitude limit the 3-layer dispersion relation $c(k)$ can be derived by solving the following equation \cite{FG}:
 \begin{equation}
K_2^2-T_1 T_2 - T_1 T_3 - T_2 T_3 = 0,
\label{3layer}
\end{equation}
where $K_j = \sqrt{N_j^2/c^2-k^2}$ and $T_j = K_j \cot(K_j H^{(3)}_j)$ ($j = 1,2,3$). Here the layer thicknesses $H^{(3)}_j$ are obtained from the piece-wise linear fits to the profiles, shown in Table \ref{table2}. Apparently, $K_j$ becomes imaginary for $k> N_j/c$, but even then the product $T_j$ remains real, ensuring the existence of periodic solutions corresponding to different modes of wave propagation. One branch of solutions can be identified with the baroclinic internal waves in the focus of the present work.}

To obtain certain data collapse in the 3-layer framework the long wave speed $c^{(3)}_0$ was used as reference to rescale the velocities to yield the nondimensional value $c''=c/c^{(3)}_0$. This $c^{(3)}_0$ was determined for each considered experiment series from (\ref{3layer}) by letting $k \to 0$. The values of $c^{(3)}_0$ are presented in Table \ref{table2}.  
In this limit the model yields $c_0^{(3)} \approx 0.96\,c_0^{(2)}$ as the long wave velocity for the stratification profiles of series \#1 and \#2, and $c_0^{(3)} \approx 0.95\,c_0^{(2)}$ for series \#3. The nondimensional values of wavenumber $k'$ are presented in the same units (i.e. $k'=k H_r$) as in the 2-layer theory for the sake of better comparison. It is to be noted that the larger $k'$, the larger the difference between the 3- and 2-layer approximation becomes. 
{The numerically obtained solutions of (\ref{3layer}) in these units are plotted in Fig.\ref{disc}a for series \#1, \#2, and \#3 (solid, dashed, and dotted curves, respectively).

The correlation plot between the theoretical $c^{(3)}_0$ and the measured (dimensional) wave speed $c$ is shown in Fig.\ref{disc}b, analogously to the 2-layer case in Fig.\ref{c_k_1}b. Apparently, the data points scatter around the $c = c^{(3)}$ line (dashed). Here the empirical fit yields $c=0.97 c^{(3)} + 0.32$ cm/s (dotted line), and the $c = c^{(3)}$ curve lies within the root mean square error of the fit (not shown). We emphasize that the deviation of a data point from the $c = c^{(3)}$ line does not appear to show any systematic connection with the wave amplitude (see color scale). This finding is somewhat surprising, since larger amplitudes are expected to necessitate nonlinear corrections in the velocities (not taken into account by the model).  

\begin{figure*}[]
\noindent\includegraphics[width=\textwidth]{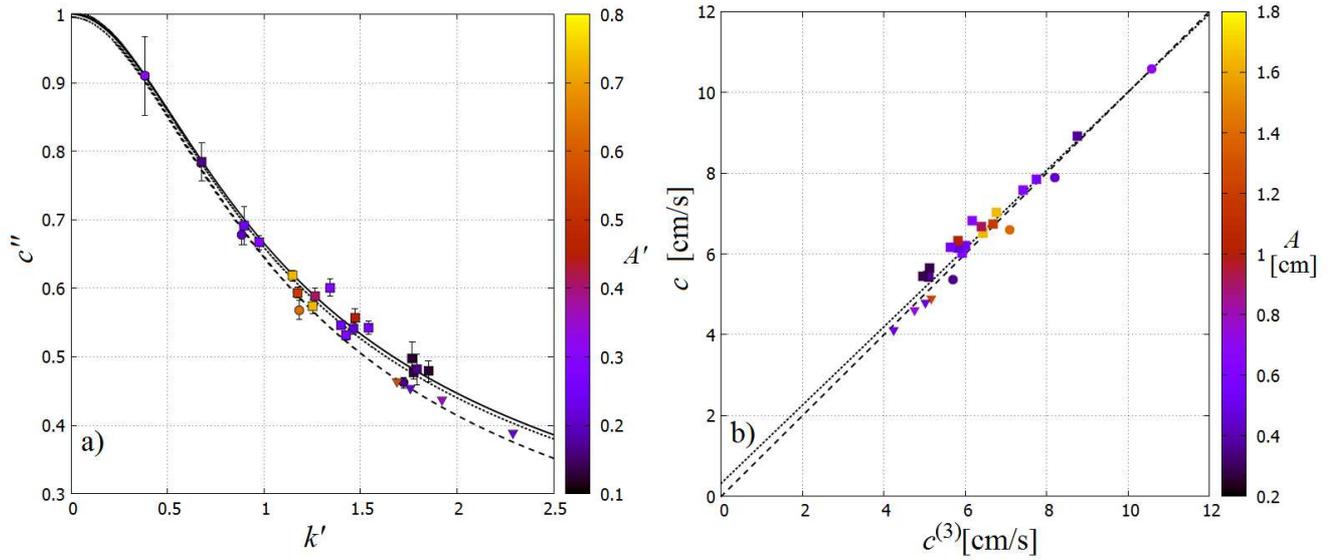}
\caption{(a) Nondimensional wave speeds $c''$ as a function of wave number $k'$ for series \#1 (red circles), series \#2 (blue squares), and series \#4 (green downward triangles). Also shown are the theoretical relations for the piece-wise linear 3-layer approximation\cite{FG} for all three experiment series (solid, dashed and dotted lines for series \#1, \#2, and \#3, respectively). The color scale indicates nondimensional amplitudes $A'$ as in Fig.\ref{c_k_1}. (b) Scatter plot of the theoretical values of 3-layer wave speed $c^{(3)}$ (horizontal axis) and the measured values $c$ (vertical axis) in dimensional units. The dashed curve represents $c=c^{(3)}$, whereas the dotted line represents the linear fit to the data. Symbol coding is as in panel (a), the color bar here represents dimensional peak-to-trough amplitudes.}
\label{disc}
\end{figure*}

\begin{figure*}[]
\centering
\noindent\includegraphics[width=0.85\textwidth]{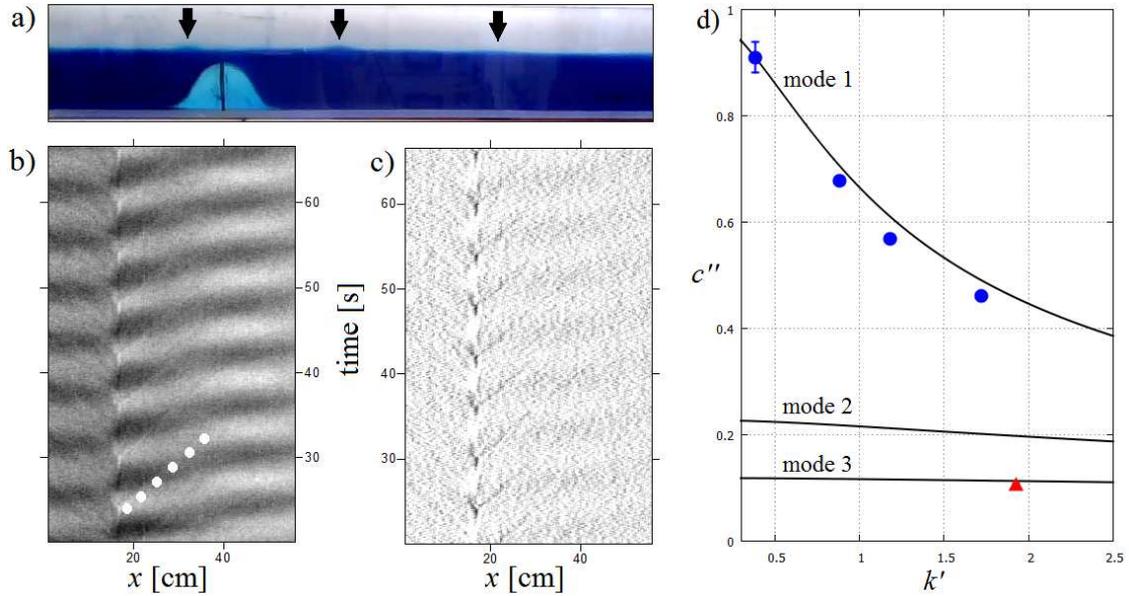}
\caption{(a) A snapshot of the experimental run at forcing frequency $\omega=1.06$ rad s$^{-1}$ in series \#1. The arrows point to the small-scale propagating waves superimposed onto the long interfacial wave. (b) The space-time plot of the experiment. The dotted line is meant to highlight the small-scale patterns which propagate along parallel trajectories. (c) The trajectories are better seen after filtering out the large-scale patterns with 21-pixel moving averaging. (d) Nondimensional wave speeds as a function of nondimensional wave number corresponding to the first three solutions (eigenmodes) of equation (\ref{3layer}) for the vertical density profile of experiment series \#1. The obtained data points for the mode 1 waves are repeated from Fig.\ref{disc}a. The speed corresponding to the small scale waves is shown by a red triangle.}
\label{pcs1}
\end{figure*}

Furthermore, it is worth mentioning that the theoretical curves shown in Fig.\ref{disc}a represent only one branch of solutions of (\ref{3layer}), namely the one of the highest wave speeds. As reported in \cite{FG} further `slow wave' modes also exist in the linear 3-layer theory which may co-exist with the dominant mode. Such a situation has probably been encountered in experiment series \#1 at forcing frequency $\omega=1.06$ rad s$^{-1}$. Fig.\ref{pcs1}a shows a typical frame from the processed video sequence with arrows pointing to certain small-scale waves. The space-time plot of the experiment is presented in Fig.\ref{pcs1}b (obtained via identical steps of data processing as the earlier plots in Fig.\ref{hovmoellers}); beside the dominant long internal waves, one can clearly see the tracks of the small disturbances as well, one of which is marked with a white dotted line. They appear to propagate at a constant velocity and remain compact until they vanish at ca. 20-30 cm horizontal distance from the obstacle. To enhance the small-scale features of the plot, we calculated the spatial (centered) moving average of the interface displacement values for each time frame using a 21-pixel (i.e. ca. 1.86 cm) window and subtracted it from the original data. The residuals are shown in Fig.\ref{pcs1}c.  

Fig.\ref{pcs1}d presents the numerical solutions of (\ref{3layer}) for the 3-layer stratification profile of experiment series \#1 (solid curves), this time showing the curves corresponding to the mode 2 and mode 3 waves as well. The data points for this particular experiment are also repeated (blue) from Fig.\ref{disc}a whereas the red triangular data point marks the average velocity calculated from the slopes of the the trajectories shown in Fig.\ref{pcs1}b. Apparently, within measurement error, the obtained wave speed is in good agreement with the calculated mode 3 velocity. Obviously, this being the single observation of this wave type in all the measurement series, no conclusive comparison can be given here with the theory, but it certainly demonstrates that interfacial wave modes of the same frequency but different speed may co-exist in the system.

\section{Discussion and conclusions}
In this experimental study we analyzed the complex interplay between surface waves and internal wave dynamics
in an enclosed laboratory tank filled-up with `quasi-2-layer' density stratified water. The forcing was imposed at the water surface by a wave maker with constant amplitude but adjustable oscillation frequency. As expected, amplified surface oscillations were detected at forcing frequencies corresponding to standing wave (seiche) modes where the integer multiples of half wavelength fits onto the length $L$ of the domain (Fig.\ref{tracersfromabove}).

Barotropic-baroclinic conversion was initiated by a bottom obstacle placed close to the midpoint of the tank, reaching up to the vicinity of the pycnocline, and thus partially blocking the flow in the bottom layer. Using dye painting and PIV observations it has been demonstrated that the combined effect of stratification and shear at the obstacle yields the formation of localized billows (shear instability). If the current above the obstacle is strong enough for these billows to develop and the period of the forcing is long enough for their growth, these disturbances can tumble over the obstacle, press down the pycnocline markedly and initiate baroclinic wave propagation. Above a certain `cutoff' frequency $\omega_*$, however, this process is inhibited: the billows do not have enough time to fully develop before the oscillating flow in the upper layer reverses. Thus the flow farther away from the obstacle stays barotropic (Figs.\ref{snapshots} and \ref{hovmoellers}). It is to be emphasized, that the value of $\omega_*$ is not a `general' threshold in itself. Its value is influenced by the current in the upper layer (that, as discussed above, is coupled to the surface waves) and the geometrical setting; if the obstacle would be farther below the interface the shear layer would not overlap with the interface, yielding a different $\omega_*$. 

The analysis of the propagating interfacial waves revealed that their wave speeds are significantly overestimated by the classic 2-layer theory (Fig.\ref{c_k_1}), but the 3-layer approximation \cite{FG} based on piece-wise linear fits to the vertical density profiles gives good agreement with the observed velocities. In concert with the expectations, higher wavenumbers yield larger deviations from the 2-layer theory (as the wavelengths get closer to the pycnocline thickness the 
2-layer approximation must break down completely). Furthermore, the 3-layer approach also predicts the existence of slower wave modes, one occurrence of which has been also observed (Fig.\ref{pcs1}).   
              
{Since the amplitudes of the internal waves observed in the experiments are not negligible compared to the vertical scale $H_r$, the question arises of whether nonlinear wave theories would give even better results.
A previous study from our laboratory \cite{boschan} addressing baroclinic wave resonance in a 2-layer configuration had found that the theory of internal cnoidal waves \cite{cnoid} was consistent with the waveforms and phase velocities measured there. These solutions of the weakly nonlinear KdV equation are based on a delicate balance between dispersion and nonlinearity, which requires that the amplitude-to-vertical scale and the square of vertical scale to wavelength ratios have to be of the same order of magnitude and both small, i.e. in our nondimensional units, $\mathcal{O}(A')=\mathcal{O}(4\pi^2/k'^2)\ll 1$ must hold. As far as the larger values of $k'$ are considered (i.e. deep and intermediate waves, where the wavelength is not much larger than the characteristic fluid depth), other nonlinear models are also available for quasi-2-layer systems \cite{Hunt1961}. What these theories all have in common is that the predicted wave speeds increase with the wave amplitudes. However, as the coloring of the data points in Figs.\ref{c_k_1} and \ref{disc} reveals, in the experiments discussed here no such correlation can be observed. The deviations from the linear 3-layer theory, demonstrated in Fig.\ref{disc}b do not show any systematic trend in terms of the peak-to-trough amplitudes. Thus, we can come to the somewhat counter-intuitive and surprising conclusion that the observed propagation of the waves in the set-up could be fairly well described by a linear theory -- at least in terms of wave speed -- and that the results do not necessitate systematic nonlinear corrections.}
  
\section*{Acknowledgments}
{The authors are grateful for Anna Koh\'{a}ri, Imre M. J\'{a}nosi and Bal\'{a}zs T\'{o}th for the crucial support.} The fruitful discussions with Tam\'{a}s T\'{e}l are also highly acknowledged.

\end{document}